\documentclass[12pt]{article}
\usepackage{epstopdf,amsfonts,amsmath,amssymb}
\usepackage{graphicx}
\DeclareGraphicsExtensions{.eps}

\newcommand\encadremath[1]{\vbox{\hrule\hbox{\vrule\kern8pt
\vbox{\kern8pt \hbox{$\displaystyle #1$}\kern8pt}
\kern8pt\vrule}\hrule}}
\def\enca#1{\vbox{\hrule\hbox{
\vrule\kern8pt\vbox{\kern8pt \hbox{$\displaystyle #1$}
\kern8pt} \kern8pt\vrule}\hrule}}

\newcommand\figureframex[3]{
\begin{figure}[bth]
\hrule\hbox{\vrule\kern8pt
\vbox{\kern8pt \vbox{
\begin{center}
{\mbox{\epsfxsize=#1.truecm\epsfbox{#2}}}
\end{center}
\caption{#3}
}\kern8pt}
\kern8pt\vrule}\hrule
\end{figure}
}
\newcommand\figureframey[3]{
\begin{figure}[bth]
\hrule\hbox{\vrule\kern8pt
\vbox{\kern8pt \vbox{
\begin{center}
{\mbox{\epsfysize=#1.truecm\epsfbox{#2}}}
\end{center}
\caption{#3}
}\kern8pt}
\kern8pt\vrule}\hrule
\end{figure}
}

\makeatletter
\@addtoreset{equation}{section}
\makeatother
\newtheorem{theorem}{Theorem}[section]

\newtheorem{remark}{Remark}[section]
\newtheorem{proposition}{Proposition}[section]
\newtheorem{lemma}{Lemma}[section]
\newtheorem{corollary}{Corollary}[section]
\newtheorem{definition}{Definition}[section]
\def\br{\begin{remark}\rm\small}
\def\er{\end{remark}}
\def\bt{\begin{theorem}}
\def\et{\end{theorem}}
\def\bd{\begin{definition}}
\def\ed{\end{definition}}
\def\bp{\begin{proposition}}
\def\ep{\end{proposition}}
\def\bl{\begin{lemma}}
\def\el{\end{lemma}}
\def\bc{\begin{corollary}}
\def\ec{\end{corollary}}
\def\beaq{\begin{eqnarray}}
\def\eeaq{\end{eqnarray}}
\newcommand{\proof}{{\noindent \bf proof: }}

\newcommand{\eproof}{$\square$}

\newcommand{\beq}{\begin{equation}}
\newcommand{\eeq}{\end{equation}}
\newcommand{\bea}{\begin{eqnarray}}
\newcommand{\eea}{\end{eqnarray}}

\newcommand{\FC}{{\mathbb F}}
\newcommand{\CC}{{\mathbb C}}
\newcommand{\ZZ}{{\mathbb Z}}

\newcommand{\curve}{{\Sigma}}
\newcommand{\ab}{{\mathfrak a}}
\newcommand{\bb}{{\mathfrak b}}

\newcommand{\genus}{{\mathfrak g}}

\newcommand{\x}{{\text{x}}}
\newcommand{\y}{{\text{y}}}

\newcommand{\Nrond}{{\stackrel{\circ}{N}}}

\newcommand{\td}{\tilde}

\usepackage[pdftex]{hyperref}
\hypersetup{colorlinks,urlcolor=magenta,citecolor=red,linkcolor=blue,filecolor=black}

\begin{document}

\sloppy

\pagestyle{empty}
\hfill IPhT-T18/052
\addtolength{\baselineskip}{0.20\baselineskip}
\begin{center}
\vspace{26pt}
{\large \bf {Notes about a combinatorial expression of the fundamental second kind differential on an algebraic curve}}
\newline
\vspace{26pt}

{B. Eynard}\hspace*{0.05cm}\footnote{\href{mailto:bertrand.eynard@ipht.fr}{bertrand.eynard@ipht.fr}}

\vspace{6pt}
Institut de Physique Th\'{e}orique de Saclay,\\
F-91191 Gif-sur-Yvette Cedex, France,
CRM, Montr\'eal, Canada.
\end{center}

\vspace{20pt}
\begin{center}
{\bf Abstract}

The zero locus of a bivariate polynomial $P(x,y)=0$ defines a compact Riemann surface $\curve$. The fundamental second kind differential is a symmetric $1\otimes 1$ form on $\curve\times \curve$ that has a double pole at coinciding points and no other pole. As its name indicates, this is one of the most important geometric objects on a Riemann surface. 
Here we give a rational expression in terms of combinatorics of the Newton's polygon of $P$, involving only integer combinations of products of coefficients of $P$. Since the expression uses only combinatorics, the coefficients are in the same field as the coefficients of $P$.

\end{center}

%-----------------------------ABSTRACT--------------------------------------
%
%Abstract

\vspace{0.5cm}

%\section*{}

%\newpage
\vspace{26pt}
\pagestyle{plain}
\setcounter{page}{1}

%*********************************************************************
%==================== ARTICLE =======================================%*********************************************************************

\section{Introduction}

%The main result is in theorem \ref{thB}.

%This is a private unpublished note, following lectures on algebraic curves, and a reply to a question asked very often by many collaborators, in particular for practical computer  implementation on symbolic mathematical languages.

\subsection{Example:}

Consider the Riemann surface $\curve$ defined as the submanifold of $\CC P^1\times \CC P^1$ of equation $y^2=x^2-1$.
A point $p$ of $\curve$ is a pair $p=(x,y)$ solution of the equation.
Let $p=(x,y)$ and $p'=(x',y')$ be 2 points of $\curve$, then the following expression (tensor product of a 1-form in $p$ by a 1-form in $p'$)
\bea\label{Bex1}
B(p,p') &=&  \frac{y y'+x x'-1}{2 \ y y' \ (x-x')^2} \  dx\otimes dx' \cr
&=& - \ \left( \frac{(y'^2-x^2+1)(y^2-x'^2+1)}{(x-x')^2(y-y')^2} +1 \right) \ \frac{dx}{2y} \otimes \frac{dx'}{2y'}
\eea
has the following properties:

$\bullet$ It is symmetric $B(p,p')=B(p',p)$.

$\bullet$ It has a double pole at $p=p'$, i.e. at $(x,y)=(x',y')$, behaving like 
\beq
B(p,p') \mathop{\sim}_{p\to p'} \frac{dx\otimes dx'}{(x-x')^2}.
\eeq

$\bullet$ It has no other poles. 
Indeed it seems to have a pole at $p=(x,y)$ and $p'=(x,-y)$ because $x'=x$, but at this point the numerator vanishes, so there is no such pole.
It seems to have a pole when $y=0$, i.e. at $x=\pm 1$, but notice that $\frac{dx}{y}=\frac{dy}{x}$ has no pole at $x=\pm 1$.
At $x\to\infty$, we have $y\sim \pm x (1+O(1/x))$, so $B$ behaves as $O(1/x^2)dx\otimes dx'$ which is analytic in a neighborhood of $\infty$, as can be seen in the coordinate $\td x=1/x$, since $\frac{dx}{x^2}=-d\td x$ has no pole at $\td x\to 0$.

$B$ is called a \textbf{fundamental second kind form} on $\curve$ (
see for instance Fay's lectures \cite{Fay} or \cite{BergSchif, Mum, KoKo, BEL, Bak1}).

\subsection{General algebraic plane curves}

In theorem \ref{thB} below, we shall generalize the second line of \eqref{Bex1} to general algebraic curves  $P(x,y)=0$ (we shall assume $P$ irreducible and sufficiently generic).
The goal is to exhibit a rational function
\beq
B((x,y),(x',y')) \in \CC(x,y,x',y') \ dx \otimes dx'
\eeq
symmetric under $(x,y)\leftrightarrow (x',y')$, and 
that has a double pole at $(x,y)=(x',y')$ with behaviour
\beq\label{Bbehavdiag}
B(p,p') \mathop{\sim}_{p\to p'} \frac{dx\otimes dx'}{(x-x')^2},
\eeq
and no other pole.
Such a form is called a \textbf{fundamental second kind form}.

The existence of fundamental 2nd kind forms is well known, $B$ can be expressed in terms of the Green function,  solution of a Dirichlet variational principle.
It can also be expressed with the Theta function and involving the Abel map $\mathfrak a$, in the form
\beq\label{Btheta}
B(p,p') = d_p \otimes d_{p'} \ \log{\Theta(\mathfrak a(p)-\mathfrak a(p')+\zeta)}
\eeq
where $\zeta$ is a zero of $\Theta$. See \cite{Fay} for details.

However, Green function, Theta function and Abel map are transcendental functions, whereas for an algebraic curve $P(x,y)=0$, there should exist, in principle, an expression of $B$ involving only rational functions of $x$ and $y$, and it is not obvious at all how to rewrite \eqref{Btheta} as a rational function of $x$ and $y$.

\smallskip

Having a rational expression for $B$ would be particularly useful if the coefficients of $P$ are in a commutative field or ring $\FC$ instead of $\mathbb C$, for example in $\mathbb Z$, in $\mathbb Q$ or a number field or a function field...
A question is whether $B$ is a rational expression with coefficients in the same ring or field, which is not apparent at all in expression \eqref{Btheta}.
A rational expression of $B$ that depends algebraically on the coefficients of $P$ is also very useful towards the study of integrable systems, see \cite{BEL}.

\smallskip
An algorithm for finding a rational expression of $B$ was developed by Baker and from Weierstrass' or Klein's works, see \cite{Bak1}. 
This algorithm simplifies for certain types of polynomials $P(x,y)$, in particular hyperelliptical cases ($\deg_y P=2$), and the $(n,s)$ curves \cite{Nak08}.

Here we shall give in theroem \ref{thB} a closed rational expression of the fundamental 2nd kind differential $B$, from the combinatorics of the Newton's polytope, valid for any polynomial $P$. This expression involves only integer combination of coefficients of $P$, and thus  elements of the same ring.
We insist that this is a very simple to use, closed and simple formula, not an algorithm. We are not aware if this formula has been known before and would be grateful to readers who could point to us where it was first derived. The formula could very well have been known to Baker or mathematicians of the 19th century.

%\medskip
%
%The main result\footnote{Although this result could have been known to Riemann, Naker, Weierstrass or other specialists of the same times, we have not found it in the literature.
%} is theorem \ref{thB}.

\section{Algebraic plane curves}
\label{sec:geom}

Let $P\in \CC[x,y]$ a bivariate irreducible polynomial  (not a product of polynomials) 
\beq
P(x,y)=\sum_{i,j} P_{i,j} x^i y^j.
\eeq
This can be generalized to Laurent polynomials, i.e. allowing negative powers of $x$ and $y$, i.e. $P\in \CC[x,1/x,y,1/y]$, so that the pairs $(i,j)$ are in $\ZZ^2$.

\bd[Newton's polygon]
The finite set $N(P)=\{(i,j)\,|\,P_{i,j}\neq 0\}\subset \mathbb Z^2$ is called the Newton's \textbf{polytope} of $P$. Its convex envelope is called the Newton's \textbf{polygon} of $P$.
We also define $\Nrond(P)$ to be the \textbf{interior} (strict) of the convex envelope of $N(P)$, and $\partial N(P)$ be the \textbf{boundary} of the convex envelope of $N(P)$.
\ed

\bd[Notation for derivatives]
We shall denote the partial derivatives of $P$ as
\beq
P_x(x,y) = \frac{\partial}{\partial x} P(x,y) 
\quad , \quad
P_y(x,y) = \frac{\partial}{\partial y} P(x,y)
\eeq
\beq
P_{x,x}(x,y) = \frac{\partial^2}{\partial x^2} P(x,y) 
\quad , \quad
P_{x,y}(x,y) = \frac{\partial^2}{\partial x \partial y} P(x,y)
\quad , \quad
P_{y,y}(x,y) = \frac{\partial^2}{\partial y^2} P(x,y)
\eeq
and so on...

\ed

\subsection{Riemann surface}

$P$ defines an algebraic plane curve 
\beq
\td\curve = \{(x,y)\in \CC P^1 \times \CC P^1 \ | \ P(x,y)=0\} \subset \CC P^1 \times \CC P^1.
\eeq
$\td\curve$ is nearly a dimension 2 submanifold of $\CC P^1\times \CC P^1$, and in fact it is the immersion in $\CC P^1 \times \CC P^1$ of a compact Riemann surface $\curve$, called its desingularization.
The immersion is realized by 2 meromorphic maps $\x:\curve\to \td\curve$ and $\y:\curve \to \td\curve$, such that
\beq
\td\curve = \{ (\x(p),\y(p)) \ | \ p\in\curve\}.
\eeq
For almost all points of $\curve$ (except a finite number), at which the normal vector $(P_x(\x(p),\y(p)),P_y(\x(p),\y(p)))\neq 0$, the map $p \mapsto (\x(p),\y(p))$ is locally invertible, and we can identify the point $p\in\curve$ with its image $(x,y)=(\x(p),\y(p))$ in $\td\curve$.
The eventual exceptional points at which the vector $(P_x(x,y),P_y(x,y))$ vanishes, are called nodal points. 
For generic $P$ there is no nodal points.

\subsection{Meromorphic functions and forms}

It is well known (and easy to prove) that the field of meromorphic functions $\mathfrak M^0(\curve)$ on $\curve$, is the field of rational functions $\CC(x,y)$ modulo $P(x,y)=0$. 
Similarly the space $\mathfrak M^1(\curve)$ of meromorphic 1-forms on $\curve$, is $\CC(x,y)dx$ modulo $P(x,y)=0$:
\beq
\mathfrak M^0(\curve) = \CC(x,y) \quad \text{modulo} \ P(x,y)=0
\eeq
\beq
\mathfrak M^1(\curve) = \CC(x,y)\ dx \qquad \text{modulo} \ P(x,y)=0 .
\eeq

\subsection{Branchpoints and double points}

At generic points $p=(x,y)\in \td\curve$, such that $P_y(x,y)\neq 0$, we can locally write $y$ as an analytic function of $x$, and we have
\beq
\frac{dy}{dx} = \frac{-P_x(x,y)}{P_y(x,y)}.
\eeq
This fails at points such that $P_y(x,y)=0$.

$\bullet$ Let $(a,y_a)$ be a point such that $P_y(a,y_a)=0$, and $P_x(a,y_a)\neq 0$, then we have
\beq
\frac{dx}{P_y(x,y)} = - \ \frac{dy}{P_x(x,y)}
\eeq
whose right hand side has no pole at $(a,y_a)$, therefore $dx$ vanishes at $(a,y_a)$.

We say that $(a,y_a)$ is a \textbf{ramification point},  and $a$ (its image by $\x$) is called a branchpoint.
We say that the ramification point is regular if $dx$ vanishes to the lowest possible order, i.e. $dx$ has a simple zero, and  $dy$ is not vanishing, and $P_y(x,y)$ has also a simple zero.

$\bullet$ Let $(b,y_b)$ be a point such that $P_y(b,y_b)=0$, and $P_x(b,y_b)= 0$. We say that $(b,y_b)$ is a \textbf{double point}.
We say that the double point is regular if $P_y$ and $P_x$ vanish to the lowest possible order, and the Hessian matrix $\begin{pmatrix} P_{x,x} & P_{x,y} \cr P_{y,x} & P_{y,y}\end{pmatrix}$ is invertible at $(b,y_b)$.
Let $\bb$ the set of all double points.

We say that $P$ is regular if all its ramification points and double points are regular.

From now on, we shall assume that $P$ is regular.
In fact, every compact Riemann surface $\curve$ can be immersed into $\CC P^1\times \CC P^1$ as the zero locus of a regular polynomial.

\subsection{Punctures}

Punctures are where $x$ and/or $y$ may tend to $0$ or $\infty$ in the locus $P(x,y)=0$.
It is well known that punctures are in 1-1 correspondence with boundaries of the Newton's polygon.

\bd[Puncture]
Let $\alpha=[(i,j)\to(i',j')]\subset \partial N(P)$ with $(i,j)\in\mathbb Z^2$ and $(i',j')\in\mathbb Z^2$, a minimal (no other integer point of $\partial N(P)$ is in $\alpha$, and thus $i-i'$ and $j-j'$ are relatively prime) boundary of the Newton's polygon, oriented such that the polygon is on its right.

The puncture $\alpha$ is such that the locus $P(x,y)=0$ has an asymptotic solution
\beq
y^{j-j'}\sim c_\alpha x^{i'-i},
\eeq
where $c_\alpha$ is a root of the following polynomial of $c$
\beq
P_\alpha(c)=\sum_{r\in \mathbb Z} P_{i-r(i'-i),j-r(j'-j)} \ c^{r} \in \CC[c,c^{-1}]
\eeq
whose degree equals the number of integer points on the boundary of $N(P)$ containing $\alpha$, and has as many roots as the number of punctures in that boundary. 
We define $c_\alpha$ the $\alpha^{\rm th}$ root (ordering the minimal oriented segments clockwise along the boundary, and the roots  by their decreasing modulus).

Since $j-j'$ and $i'-i$ are relatively prime, there exist $\beta_\alpha$ and $\td\beta_\alpha$ such that
\beq
\beta_\alpha(j'-j)+\td\beta_\alpha(i-i')=1.
\eeq
We define
\beq\label{defzlocalcoordalpha1}
z=x^{\beta_\alpha} y^{\td\beta_\alpha}.
\eeq
$z$ is a  local coordinate in the neighborhood of the puncture $\alpha$.
In that neighborhood we have
\beq\label{xyzlocalcoordalpha1}
x\sim c_\alpha^{\td\beta_\alpha} \ z^{j'-j}
\qquad , \qquad
y\sim c_\alpha^{-\beta_\alpha} \ z^{i-i'}.
\eeq
We see that indeed near that point, $x$ and/or $y$ have a pole or a zero, i.e. this is a puncture.
\ed

Moreover, all punctures, i.e. points $\alpha\in\curve$ where $x$ and/or $y$ have a pole or zero, are of that type.
Indeed, let us assume that there is a point $\alpha\in\curve$, and a local coordinate $z$ near $\alpha$, vanishing at $\alpha$, and such that
\beq\label{defzlocalcoordalpha1}
x\sim c\ z^{-p}
\qquad , \qquad
y\sim \td c \ z^{-q},
\eeq
where $c\neq 0$ and $\td c\neq 0$, and 
where $p$ and $q$ are integers not both vanishing, i.e. $x$ has a pole if $p>0$ or zero if $p<0$ or none if $p=0$, and $y$ has a pole if $q>0$ or zero if $q<0$ or none if $q=0$.
Let 
\beq\label{defmpq}
m_{p,q} = \max_{(i,j)\in N(P)}  pi+qj. 
\eeq
Define the line $D_{p,q}$ of equation $pi+qj=m_{p,q}$
\beq
D_{p,q} = \{(i,j)\in \mathbb R^2 \ | \ pi+qj=m_{p,q} \}.
\eeq
By definition $D_{p,q}\cap N(P)\neq 0$, and due to maximality \eqref{defmpq} the Newton's polytope lies to the right of that line (oriented as the vector $(q,-p)$).

We have
\beq
0  =  P(x,y) = \sum_{(i,j)\in N(P)} P_{i,j} x^i y^j \sim  z^{m_{p,q}} \ \sum_{(i,j)\in N(P)\cap  D_{p,q}} P_{i,j} c^i \td c^j  + o(z^{m_{p,q}}),
\eeq
whose right hand side can't be zero if $D_{p,q}$ would contain only one point of $N(P)$, therefore $D_{p,q}$ must contain at least 2 points $(i,j)$ and $(i',j')$.
This implies that
\beq
p(i-i')=q(j'-j),
\eeq
and also this implies that the line $D_{p,q}$ is tangent to the convex envelope of $N(P)$.
This implies that $\alpha$ is a puncture as defined above.

\subsection{Holomorphic forms}

Let ${\mathcal O}^1(\curve)\subset  {\mathcal M}^1(\curve)$ the space of "holomorphic forms": forms that have no poles at all.
It is well known that if $P$ is regular and  has no double point, the space of holomorphic forms on $\curve$ is given by the interior of Newton's polygon
\bt[Holomorphic forms]
\bea
\mathcal O^1(\curve) 
%&=& \frac{dx}{xy \ P_y(x,y)} \ \CC^{\Nrond(P)} \cr
&=& \left\{ \,\frac{Q(x,y)\,dx}{P_y(x,y)} \, \,\, \Big|\,\,Q(x,y)=\sum_{(u,v)\in\Nrond(P)}\,Q_{u,v}\,x^{u-1}\,y^{v-1} \,\, , \,\, Q_{u,v}\in \CC\right\}.
\eea
We have
\beq
\mathcal O^1(\curve) \sim \CC^{\Nrond(P)}
\qquad , \qquad
\dim \mathcal O^1(\curve) =|\Nrond(P)|= \genus.
\eeq
\et

\proof
This is a well known theorem. Let us recall the proof because it will be useful for understanding the similar proof for the fundamental 2nd kind form.

The ratio $\frac{dx}{P_y(x,y)}= - \ \frac{dy}{P_x(x,y)} $ has no pole at the zeros of $P_y(x,y)$ i.e. at ramification points.
The only places where $\frac{x^{u-1} y^{v-1} \ dx}{P_y(x,y)}$ could have poles is at punctures.
Let $\alpha$ a puncture, and let $z$ a local coordinate as in \eqref{defzlocalcoordalpha1}. In the neighborhood of $\alpha$ we have
\bea
P_y(x,y) 
&\sim & \sum_{(i'',j'')\in D_{p,q}\cap N(P)} P_{i'',j''} j''  c^{i''}\td c^{j''} z^{-pi''-q(j''-1)} \cr
&\sim & z^{-m_{p,q}+q} \ \sum_{(i'',j'')\in D_{p,q}\cap N(P)} P_{i'',j''} j''  c^{i''}\td c^{j''}  \cr
&\sim & C \ z^{-m_{p,q}+q} 
\qquad , \quad C\neq 0.
\eea
Moreover we have $dx = -p c \ z^{-p-1} dz$, therefore
\beq
\frac{x^{u-1} y^{v-1} \ dx}{P_y(x,y)} \sim O(z^{-up-vq+m_{p,q}-1}) dz.
\eeq
It has no pole at $\alpha$, i.e. at $z=0$ if and only if $up+vq-m_{p,q}<0$, i.e. the point $(u,v)$ lies strictly to the right of the line $D_{p,q}$.
If this is true for all punctures, then $(u,v)$ must lie to the right of every tangent of the convex envelope, i.e. $(u,v)\in \Nrond(P)$.
Reciprocally, if $(u,v)\in \Nrond(P)$, then the form $\frac{x^{u-1} y^{v-1} \ dx}{P_y(x,y)}$ has no pole at all, it is holomorphic.
\eproof

In the proof we have seen that more generally, for $(u,v)\in\mathbb Z^2$, the form
\beq
\frac{x^{u-1}y^{v-1}\ dx}{P_y(x,y)}
\eeq
 has 
\begin{itemize}

\item no pole at $\alpha$ if $(u,v)$ is on the right of the line $D_{p,q}$,

\item a simple pole at $\alpha$ if $(u,v)$ is on the line $D_{p,q}$,

\item a pole at $\alpha$ of degree  $1+up+vq-m_{p,q}\geq 2$ if $(u,v)$ is on the left of $D_{p,q}$.

\end{itemize}

When $P$ has double points, we would like the ratio $\frac{Q(x,y)dx}{P_y(x,y)}$ to have no pole at double points (which are zeros of $P_y(x,y)$ but not of $dx$), i.e. that $Q(b,y_b)=0$ for each double point, in other words
\bp[Holomorphic forms, double-points]
If $P$ is regular and  has double points, we have
\beq
\mathcal O^1(\curve) = \operatorname{Ker}\left(
\begin{array}{l}
\CC^{\Nrond(P)}\mapsto \CC^{\bb} \cr
Q\mapsto \{Q(b,y_b)\}_{b\in \bb}
\end{array}\right)
\eeq
i.e.
\beq
\mathcal O^1(\curve) 
= \left\{ \,\frac{Q(x,y)\,dx}{P_y(x,y)} \, \,\, \Big|\,\,Q(x,y)=\sum_{(i,j)\in\Nrond(P)}\,Q_{i,j}\,x^{i-1}\,y^{j-1} \,\, , \forall \ b \in \bb \  \ Q(b,y_b)=0 \ \right\}.
\eeq
We have the exact sequence
\beq
0 \to \mathcal O^1(\curve) \to \CC^{\Nrond(P)}
 \to \CC^{\bb} \to 0.
\eeq
We have
\beq
\dim \mathcal O^1(\curve) = \genus =|\Nrond(P)|-|\bb|.
\eeq
\ep

In appendix A, we show that if the coefficients of $P$ are in a commutative field $\FC$ of characteristic 0, then there is a basis of $\mathcal O^1(\curve)$ with polynomials $Q\in \FC[x,y]$ the same field as $P$. This is not obvious, because if $\FC$ is not algebraically complete, the double points are in an algebraic extension of $\FC$, and a priori the coefficients $Q_{i,j}$ would only be in that extension.

\section{Fundamental 2nd kind form}

Let us look for a fundamental 2nd kind form as follows
\beq\label{Bcandidate}
B(x,y;x',y') = - \ \frac{\frac{P(x,y')P(x',y)}{(x-x')^2(y-y')^2} - Q(x,y;x',y')}{P_y(x,y) P_y(x',y')} \ dx \ dx' 
\eeq
with $Q\in \mathbb C[x,y,x',y']$ a polynomial of all its variables (resp. $Q\in \mathbb C[x,1/x,y,1/y,x',1/x',y',1/y']$ if $P$ is a Laurent polynomial), symmetric under $(x,y)\leftrightarrow (x',y')$, chosen so that $B(x,y)$ has a pole only at $(x,y)=(x',y')$ on the curve $\curve$, in particular, $Q$ must be chosen so that $B$ has no pole at punctures.

Remark that \eqref{Bcandidate} has a double pole at $(x,y)=(x',y')$, indeed near that point 
\bea
P(x,y') = P(x,y')-P(x,y) \sim (y'-y) P_y(x,y) \cr
P(x',y) = P(x',y)-P(x',y') \sim (y-y') P_y(x,y),
\eea
which implies behavior \eqref{Bbehavdiag}.

Remark that there is no pole at points $y=y'$ with $x\neq x'$, indeed at such points
\beq
\frac{P(x,y')P(x',y)}{(x-x')^2(y-y')^2} \sim \frac{- \ P_y(x,y)P_y(x',y)}{(x-x')^2}
\eeq
which has no pole if $x\neq x'$.
Similarly there is no pole at points $x=x'$ with $y\neq y'$.

\subsection{Regular polynomial, without double points}

Let us first consider the case where $P$ is regular and has no double points.
In that case we know that $\frac{dx}{P_y(x,y)}$ has no pole at ramification points.
The only possible other poles of \eqref{Bcandidate} could be at punctures.
We have to choose $Q$ in order to kill them.

The following is the main theorem of this article:

\bt[Fundamental 2nd kind form]\label{thB}
the following $1\otimes 1$ form
\beq\label{defBarithm}
B((x,y);(x',y')) = - \ \frac{\frac{P(x,y')P(x',y)}{(x-x')^2(y-y')^2} - Q(x,y;x',y')}{P_y(x,y) P_y(x',y')} \ dx \ dx' 
\eeq
with
\bea\label{defQBarithm}
Q(x,y;x',y')
&=& \sum_{(i,j)\in \mathcal N} \sum_{(i',j')\in \mathcal N} P_{i,j} P_{i',j'}
\sum_{(u,v)\in \ZZ^2\cap \text{ triangle }(i,j),(i',j'),(i,j')}
|u-i| \ | v-j'| \cr
&& \Big( \delta_{(u,v)\notin {\stackrel{\circ}{\mathcal N}} \cup [(i,j),(i',j')] } \ \ x^{u-1}y^{v-1} x'^{i+i'-u-1}y'^{j+j'-v-1} \cr
&& +\delta_{(u,v)\notin {\stackrel{\circ}{\mathcal N}} \text{ and } (i+i'-u,j+j'-v)\in {\stackrel{\circ}{\mathcal N}} } \ \ x'^{u-1}y'^{v-1} x^{i+i'-u-1}y^{j+j'-v-1}  \cr
&& + \frac12\ \delta_{(u,v)\in  [(i,j),(i',j')]} \ \ x^{u-1}y^{v-1} x'^{i+i'-u-1}y'^{j+j'-v-1}  \Big) \ .
\eea
is a fundamental 2nd kind form.
Here $\delta_C$ means 1 if the condition $C$ is fulfilled, and 0 if not.

\et

\textbf{Remark:} every term in the last line appears twice in the sum, because it is symmetric in $(i,j)\leftrightarrow(i',j')$. We could remove the $\frac12$ factor, by summing only over distinct pairs $\{(i,j),(i',j')\}$.

It is obvious from expression \eqref{defQBarithm}, that if the coefficients $P_{i,j}$ belong to a commutative ring $\FC$, then the coefficients of $Q$ are in the same ring.
\medskip

\proof
We have to show that $Q$ is symmetric, and $B$ has no pole at punctures.

Let us consider a puncture $\alpha$ where $x \sim c z^{-p} $ and $y\sim \td  z^{-q} $, with both $p>0$ and $q>0$ (the other cases, can be obtained by changing $x\to 1/x$ and/or $y\to 1/y$, and remarking that expression \eqref{defBarithm} is unchanged under these changes. The cases where $p=0$ or $q=0$ can be done similarly).
Near $(x,y)\to \alpha$ we have $x\to\infty$ and $y\to\infty$, and thus
\bea
\frac{P(x,y')P(x',y)}{(x-x')^2(y-y')^2} 
&\sim & \sum_{(i,j)\in \mathcal N} \sum_{(i',j')\in\mathcal N} \sum_{k\geq 1} \sum_{l\geq 1} P_{i,j} P_{i',j'} \ k l \ x^{i-k-1}y'^{j+l-1} x'^{i'+k-1}y^{j'-l-1} \cr
&\sim & \sum_{(u,v)\in \ZZ^2} x^{u-1} y^{v-1} \ \Big( \sum_{(i,j)\in \mathcal N} \sum_{(i',j')\in\mathcal N} \sum_{k\geq 1} \sum_{l\geq 1} P_{i,j} P_{i',j'} \ k l \cr
&& \ \delta_{u,i-k}\delta_{v,j'-l} x'^{i'+k-1}y'^{j+l-1} \Big) 
\eea
where the last bracket contains in fact a finite sum.
All the monomials such that $(u,v)\notin \stackrel{\circ}{\mathcal N} $ and $(u,v)$ is above the line $D_{p,q}$, gives a pole at $\alpha$, and must be compensated by a term in $Q$. 

Let us consider such an $(u,v)$ monomial.
Notice that $(u,v)$ above $D_{p,q}$ implies that
$u=i-k\geq i'$ and $v=j'-l\geq j$, which implies in particular that this can occur only if $i>i'$ and $j'>j$.
Moreover, since all the segment $[(i,j),(i',j')]$ is contained in the Newton's polygon and is not left of $D_{p,q}$, we see that $(u,v)$ must belong to the triangle $((i,j),(i',j'),(i',j))$.

Consider the point $(u',v')=(i'+k,j+l)=(i+i'-u,j+j'-v)$, which is the symmetric of $(u,v)$ with respect to the middle of $[(i,j),(i',j')]$.

\smallskip
So, let us thus assume that $i>i'$ and $j'>j$ and $(u,v)$ belongs to the triangle $((i,j),(i',j'),(i',j))$, and let us consider different cases:

$\bullet$ $(u,v)\notin [(i,j),(i',j')] $, i,e. it is strictly in the interior of the triangle.
If $(u,v)\notin \stackrel{\circ}{\mathcal N}$, then the monomial $P_{i,j} P_{i',j'} k l x^{u-1}y^{v-1} x'^{u'-1} y'^{v'-1}$ should appear in $Q$ and is indeed the first term in \eqref{defQBarithm}.
Notice that in that case the point $(u',v')$ can't be above $D_{p,q}$.
There are then 2 sub-cases:

$\bullet\bullet$ $(u',v')\in \stackrel{\circ}{\mathcal N}$, then we can add to $Q$ a monomial proportional to $x^{u'-1}y^{v'-1}$ without adding a pole to $B$. In particular we can add
\beq
P_{i,j}P_{i',j'} \ k l \ x^{u'-1}y^{v'-1} x'^{u-1} y'^{v-1}
\eeq
which is the second term in \eqref{defQBarithm}.
It is the term needed to make $Q$ symmetric under the exchange $(x,y)\leftrightarrow (x',y')$, and that doesn't appear anywhere else.

$\bullet\bullet$ $(u',v')\notin \stackrel{\circ}{\mathcal N}$.
Notice that since $(u,v)\notin [(i,j),(i',j')] $, we also have $(u',v')\notin [(i,j),(i',j')] $.
Moreover, if $(u',v')\notin \stackrel{\circ}{\mathcal N}$, this implies that $(u',v')$ is in the triangle $(i',j'),(i,j),(i',j)$ and thus below some tangent $D_{p',q'}$ with $p'\leq 0$ and $q'\leq 0$.
This means that the monomial $P_{i,j}P_{i',j'} kl x^{u'-1}y^{v'-1} x^{u-1}y^{v-1}$ will appear in $Q$ in the contribution with $(i,j)\leftrightarrow(i',j')$.
This guarantees that this contribution to $Q$ is symmetric in $(x,y)\leftrightarrow (x',y')$.

$\bullet$ $(u,v)\in [(i,j),(i',j')] $.
This implies that $(u',v')\in [(i,j),(i',j')] $ as well.
Remarking that if $(u,v)\in [(i,j),(i',j')] $, we have $kl=(i-u)(j'-v)=(u-i')(v-j) $, we have
\beq
P_{i,j} P_{i',j'} (i-u) (j'-v) x^{u-1}y^{v-1} x'^{u'-1} y'^{v'-1}
=
P_{i',j'} P_{i,j} (i'-u) (j-v) x^{u-1}y^{v-1} x'^{u'-1} y'^{v'-1}
\eeq
i.e. this monomial appears twice in the sum \eqref{defQBarithm} because it also appears in the term $(i,j)\leftrightarrow(i',j')$, and this is why it has to be multiplied by $\frac12$.

Also, if $(u,v)\in [(i,j),(i',j')] $, we have $kl=(i-u)(j'-v)=(i-u')(j'-v')$, the monomial $P_{i,j}P_{i',j'} kl \ x^{u'-1}y^{v'-1} x^{u-1}y^{v-1}$ also appears in \eqref{defQBarithm}, and this guarantees that this contribution to $Q$ is symmetric under $(x,y)\leftrightarrow (x',y')$.

Also, if $(u,v)\in \stackrel{\circ}{\mathcal N}\cap [(i,j),(i',j')] $, this implies that $(u',v')\in \stackrel{\circ}{\mathcal N}\cap [(i,j),(i',j')] $, and thus this monomial and its symmetric under $(x,y)\leftrightarrow(x',y')$ are both inside Newton's polygon, so don't contribute to poles of $B$.

\smallskip

Eventually we have shown that the polynomial of \eqref{defQBarithm} is symmetric under $(x,y)\leftrightarrow(x',y')$, and up to monomials inside $\stackrel{\circ}{\mathcal N}$, it compensates all the terms of $\frac{P(x,y')P(x',y)}{(x-x')^2(y-y')^2} $ that could possibly diverge at punctures.

This concludes the proof.
\eproof

\bc[Shifted fundamental 2nd kind form]
Every fundamental 2nd kind form can be written as
\beq
B_\kappa(x,y;x',y') = \frac{\left(\frac{P(x,y')-P(x',y)}{(x-x')(y-y')}\right)^2 - Q_\kappa(x,y;x',y')}{4 \ P_y(x,y) P_y(x',y')} \ dx \ dx' .
\eeq
where $\kappa$ is a symmetric matrix in $\CC^{\Nrond\times \Nrond}$, and
\beq
Q_\kappa(x,y;x',y') = Q(x,y;x',y') - \sum_{(i,j),(i',j')\in \Nrond\times \Nrond} \kappa_{(i,j),(i',j')} \ x^{i-1}y^{j-1} x'^{i'-1} y'^{j'-1}.
\eeq
\ec

\proof
Indeed one can add to $B$ any symmetric bilinear combination of holomorphic forms  $\mathcal O^1(\curve)\otimes^{\text{sym}} \mathcal O^1(\curve)$.

Conversely, if $\td B$ is another fundamental 2nd kind form, then $\td B-B$ must have no pole at all, it must be a 
symmetric bilinear combination of holomorphic forms  $\mathcal O^1(\curve)\otimes^{\text{sym}} \mathcal O^1(\curve)$.
\eproof

\subsection{Double points}

\bt[Fundamental 2nd kind form: double points case]
If $P$ has double points,  then the following is a fundamental 2nd kind form:
\beq
B(x,y;x',y') = - \ \frac{\frac{P(x,y')P(x',y)}{(x-x')^2(y-y')^2} - Q(x,y;x',y') - \td Q(x,y;x',y')}{P_y(x,y) P_y(x',y')} \ dx \ dx' 
\eeq
with $Q$ defined in \eqref{defQBarithm}, and $xyx'y'\td Q \in \CC^{\Nrond(P)} \stackrel{\text{sym}}{\otimes} \CC^{\Nrond(P)}$ defined as follows:
let us choose $I\subset \Nrond(P)$ such that $|I|=|\bb|$, we recall that the matrix $B_I$ defined in lemma \ref{lemmabb} is invertible.
We define for every $(i,j),(i',j')\in I\times I$:
\bea\label{deftdQthB}
\td Q_{i,j;i',j'}
%&=& 0 \qquad \text{if} \ (i,j),(i',j')\notin I\times I \cr
&=& -\sum_{b\neq b'\in \bb\times \bb} (B_I^{-1})_{(i,j),b} (B_I^{-1})_{(i',j'),b'}\ \left(
Q(b,y_b;b',y_{b'}) - \frac{P(b,y_{b'})P(b',y_b)}{(b-b')^2(y_b-y_{b'})^2} \right)\cr
&& - \sum_{b\in \bb} (B_I^{-1})_{(i,j),b} (B_I^{-1})_{(i',j'),b}\ \left(
Q(b,y_b;b,y_{b}) - \gamma_b\right) 
%\qquad \text{if} \ (i,j),(i',j')\in I \times I \cr
\eea
where $\gamma_b = P_{x,y}(b,y_b)^2 - P_{x,x}(b,y_b) P_{y,y}(b,y_b)$, and finally
\beq
\td Q(x,y;x',y') = \sum_{(i,j),(i',j')\in I\times I} \td Q_{i,j;i',j'} \ x^{i-1}y^{j-1} x'^{i'-1} y'^{j'-1}.
\eeq

\et

\proof
Since $\td Q$ has its coefficients in $I\subset \Nrond(P)$ it does not create poles at punctures, and by construction it kills poles at double points.
\eproof

%\subsubsection{Examples}
%
%Case $\deg_y P=2$:
%\beq
%P(x,y)=y^2-U(x)
%\eeq
%with $\deg U=d$.
%We have
%\beq
%\Nrond=\left\{(a,1) \ | \ 1\leq a < \frac{d}{2} \right\}.
%\eeq
%We have
%\bea
%Q(x,y;x',y')
%&=& \sum_{i'=2}^{\deg U} P_{0,2}P_{i',0} \sum_{k=1}^{i'-1} k \ \chi(i'-k,1;k,1) x^{i'-k-1}x'^{k-1} \cr
%&& + \sum_{i=2}^{\deg U} P_{i,0}P_{0,2} \sum_{k=1}^{i-1}  \ \chi(k,1;i-k,1) x^{k-1}x'^{i-k-1} \cr
%\eea
%% i'-k<d/2 et k>d/2
%% i'-k>d/2 et k<d/2
%
%\beq
%Q(x,y;x',y')
%= \frac{d}{2} \ P_{0,2}P_{d,0}  x^{\frac{d}{2}-1}x'^{\frac{d}{2}-1}  + \sum_{j>d/2}^{d-1} \sum_{k=1}^{d-j}
% P_{0,2}P_{j+k,0}  k \  (x^{j-1}x'^{k-1}+x'^{j-1}x^{k-1}) 
%\eeq
%
%
%\subsubsection{Example}
%
%\beq
%P(x,y)=V(y)-U(x)
%\eeq
%with $\deg U=d$, $\deg V=d'$.
%The Newton's polygon is a triangle:
%\beq
%\Nrond=\left\{(a,b) \ | \ 1\leq a, \ 1\leq b, \ d'a+db < dd' \right\}.
%\eeq
%We get
%\bea
%Q_{u,v;u',v'} = P_{u+u',0} P_{0,v+v'} u' v'
%\eea
%
%

\subsection{Hyperelliptical case}

Consider the case $P(x,y)=y^2-P(x)$, with $P(x)\in\mathbb C[x]$ a polynomial of even degree, whose zeros are all distinct (thus no double points).
Theorem \ref{thB} gives
\bea
B(x,y;x',y') 
&=& \frac{dx}{P_y(x,y)}\frac{dx'}{P_y(x',y')} \ \Big( \frac{2yy'}{(x-x')^2} + \frac{P(x)+P(x')}{(x-x')^2} \cr
& & - \sum_i i P_{2i} x^{i-1}x'^{i-1}  \cr
&& - \sum_{i>\frac12 \deg P} P_i \sum_{\frac12 i<u<i} (i-u) (x^{u-1}x'^{i-u-1}+x^{i-u-1}x'^{u-1} ) \Big) \cr
\eea

In that case we can further simplify the expression of $B$. See \cite{BEL}.

\bt[Hyperellitical curves]
Let $U(x)=(\sqrt{P(x)})_+$ be the polynomial part (expanding near $\infty$) of its square-root, and let $V(x)=P(x)-U(x)^2$.
Then
\beq
B((x,y);(x',y')) = \frac{y y' + U(x) U(x')+\frac12 V(x)+\frac12 V(x')}{2y y' (x-x')^2} \ dx \ dx'.
\eeq
is a fundamental 2nd kind form.
\et

\proof
The symmetry $(x,y)\leftrightarrow (x',y')$ is obvious.
The double pole at $(x,y)=(x',y')$ is obvious.
The absence of pole at $(x',y')=(x,-y)$ is also obvious, and the absence of poles at the zeros of $y$ as well.
The only possible poles could come from the 2 punctures  $x\to\infty$, and $y\sim \pm U(x)$. Notice that $\deg V<\deg U$.
Consider the $+$ sign (the minus sign is similar).
At this puncture we have
\beq
y = U(x) + \frac{V(x)}{2U(x)} + O(1/x^2),
\eeq
therefore
\beq
\frac{y y' + U(x) U(x')+\frac12 V(x)+\frac12 V(x')}{2y y'} \sim \frac{y'+U(x')}{2y'} + O(1/x)
\eeq
and thus
\beq
B((x,y);(x',y'))  \sim \frac{dx}{x^2} \ \left(\frac{y'+U(x')}{2y'} + O(1/x)\right) dx'
\eeq
which is regular at $x\to\infty$.
\eproof

\subsection{$(n,s)$ curves}

$(n,s)$ curves are those whose Newton's polygon is a triangle:
\beq
P(x,y)=y^n-x^s-\sum_{i\geq 0,j\geq 0, ni+js<ns} P_{i,j} x^i y^j.
\eeq
We assume that $n$ and $s$ are relatively prime.

In that case we get
\bea
Q(x,y;x',y') 
&=& -\sum_{k>0,l>0,kn+ls<ns} (s-k)(n-l) (x^{s-k-1}y^{n-l-1}x'^{k-1}y'^{l-1}\cr
&& \qquad \qquad +x^{k-1}y^{l-1}x'^{s-k-1}y'^{n-l-1}) \cr
&& + \sum_{0<u<s,0<v<n, un+vs>ns}\qquad \sum_{u'>0,v'>0, u'n+v's<ns}  \cr
&& \sum_{k=1}^{\min(u',s-u)-1}\sum_{l=1}^{\min(v',n-v)-1} k l \ P_{u+k,v'-l} P_{u'-k,v+l} (x^{u-1}y^{v-1}x'^{u'-1}y'^{v'-1}\cr
&& \qquad \qquad +x^{u'-1}y^{v'-1}x'^{u-1}y'^{v-1})\cr
&& + \sum_{(i,j)} \qquad \sum_{(u,v)\in [(i,j),(s,0)]}  (u-i) v \ P_{i,j} \  x^{u-1}y^{v-1} x'^{i+s-u-1} y'^{j-v-1} \cr
&& - \sum_{(i,j)} \qquad \sum_{(u,v)\in [(i,j),(0,n)]}  (i-u) (n-v) \ P_{i,j} \  x^{u-1}y^{v-1} x'^{i-u-1} y'^{j+n-v-1} \cr
&& + \sum_{\text{pairs}\{(i,j),(i',j')\}} \qquad \sum_{(u,v)\in [(i,j),(i',j')]} \cr
&& |u-i| |v-j'| \ P_{i,j}P_{i',j'} \  x^{u-1}y^{v-1} x'^{i+i'-u-1} y'^{j+j'-v-1}
\eea
Notice that the last 3 lines are in fact in $\Nrond(P)$ and could be removed, we could keep only
\bea
&& -\sum_{k>0,l>0,kn+ls<ns} (s-k)(n-l) (x^{s-k-1}y^{n-l-1}x'^{k-1}y'^{l-1}\cr
&& \qquad \qquad +x^{k-1}y^{l-1}x'^{s-k-1}y'^{n-l-1}) \cr
&& + \sum_{0<u<s,0<v<n, un+vs>ns}\qquad \sum_{u'>0,v'>0, u'n+v's<ns}  \cr
&& \sum_{k=1}^{\min(u',s-u)-1}\sum_{l=1}^{\min(v',n-v)-1} k l \ P_{u+k,v'-l} P_{u'-k,v+l} (x^{u-1}y^{v-1}x'^{u'-1}y'^{v'-1}\cr
&& \qquad \qquad +x^{u'-1}y^{v'-1}x'^{u-1}y'^{v-1})\cr
\eea

See \cite{Nak08}.

Example $(n,s)=(4,3)$:
\bea
Q(x,y;x',y') 
&=& -  (xy^{2}+x'y'^2) - 2 (xyy'+yx'y')-2(y^2x'+xy'^2)   + P_{0,3}P_{2,0}(y+y') \cr
&& - 2 P_{2,0} yy' + 2 P_{0,3} (xx'+yy') + P_{1,2} xx'  + P_{0,3}P_{2,1} yy' + P_{0,2}P_{2,0} \cr
\eea
where all terms in the 2nd line could be discarded.

\section{Other fields}

Let us assume that all coefficients of $P$ are in a commutative field $\FC$ rather than $\CC$, for example $\mathbb Q$ or a number field or a function field.
Are the coefficients of $Q$ (and $\td Q$) in $B$ in the same field ?

If we have no double points, this is rather obvious, and we can replace commutative field $\FC$ by commutative ring: the coefficients of $Q$ are integer bylinear combinations of coefficients of $P$ so they are always in the same ring.

If there are double points, it is more subtle to see that the polynomial $\td Q$ in \eqref{deftdQthB} has its coefficients in $\FC$.
Let us study this question now in greater details.

\subsection{Branchpoints and double points}

Assume that $\FC$ is a commutative field.

\bd[Discriminant]
We define the $x$-discriminants:
\beq
D(x) = {\rm Resultant}_y(P(x,y) , P_y(x,y)\, ) \,\, \in \FC[x].
\eeq
\beq
\td D(x) = {\rm Resultant}_y(P(x,y) , P_x(x,y)\, ) \,\, \in \FC[x].
\eeq
And we define the discriminant $\Delta$ as the discriminant of $D(x)$, i.e. the resultant of $D(x)$ and $D'(x)$.
\beq
\Delta = {\rm Discriminant}(D)= {\rm Resultant}(D,D'),
\qquad \quad \Delta\in \FC.
\eeq

\ed

\bd[Branchpoints]

Let us consider the set of roots of $D(x)$ in an algebraic closure of $\FC$.

$\bullet$ If $a$ is a root of $D(x)$ and not a root of $\td D(x)$, we say that $a$ is a branchpoint. 
Let 
\beq
\ab = \{a_1,\dots,a_{|\ab|}\}
\eeq
the set of all branchpoints.
We shall consider the field extension
\beq
\CC[\ab]=\CC[a_1,\dots,a_{|\ab|}].
\eeq

$\bullet$ If $b$ is a root of $D(x)$ and a root of $\td D(x)$, we say that $b$ is a double point. 
Let 
\beq
\bb = \{b_1,\dots,b_{|\bb|}\}
\eeq
the set of all double points.

\ed

\bl\label{lemsymbpFC}
The  symmetric polynomials (over $\FC$) of $b_1,\dots,b_{|\bb|}$ are in $\FC$.
The symmetric polynomials of $a_1,\dots,a_{|\ab|}$ are in $\FC$.

\el

\proof
In appendix \ref{appprooflemmabsym}
\eproof

\bt
If $\FC$ is a commutative field of characteristic 0, then the coefficients of $\td Q$ in \eqref{deftdQthB} are in $\FC$.
\et

\proof
In appendix \ref{approoflemmabb}.
\eproof

\section*{Conclusion}

The formula has a nice combinatorial graphical interpretation: it is a sum over oriented pairs of distinct points in the Newton's polygon, each pair $(i,j),(i',j')$ defines a triangle $(i,j),(i',j'),(i,j')$, and then we sum over all points $(u,v)$ in that triangle, yielding a monomial $P_{i,j}P_{i',j'} |u-i||v-j'| \ x^{u-1}y^{v-1}x'^{i+i'-u-1} y'^{j+j'-v-1}$, depending on whether the point $(u,v)$ is on the segment $[(i,j),(i',j')]$, and depending whether the point $(u,v)$ is inside $\Nrond(P)$ or outside, and also depending on whether the symmetric (with respect to the middle of $[(i,j),(i',j')]$) of $(u,v)$ is inside or outside $\Nrond(P)$.

The factor $|u-i||v-j'|$ is the area of the rectangle of corners $(u,v),(i,j')$.

This is extremely easy to implement.

\section*{Acknowledgments}

This work was supported by the ERC Starting Grant no. 335739 ``Quantum fields and knot homologies'' funded by the European Research Council under the European Union's Seventh Framework Programme. 
It is also partly supported by the ANR grant Quantact : ANR-16-CE40-0017.
I wish to thank the Centre de Recherches Math\'ematiques de Montr\'eal, the FQRNT grant from the Qu\'ebec government.
I also thank G. Borot, V. Bouchard, D. Korotin for discussions on that topic.

\appendix{}

\section{Lemma for double points}
\label{approoflemmabb}

\bl\label{lemmabb}
The matrix $B=(b^i,y_{b}^j)_{b\in \bb, \ (i,j)\in \Nrond(P)}$ has rank $|\bb|$.
Let $I\subset \Nrond(P)$ such that $|I|=|\bb|$.
The matrix $B_I=(b^i,y_{b}^j)_{b\in \bb, \ (i,j)\in I}$  is invertible.
Let $\bar I=\Nrond(P)-I$ the complement of $I$.

Then, for each $(i',j')\in \bar I$, the following
\beq\label{defQijlemma}
Q^{(i',j')}_{i,j} = 
\left\{\begin{array}{l}
\det_{b\in \bb,  (u,v)\in I-(i,j)+(i',j')}\left(b^u y_b^v \right)/\det_{b\in \bb,  (u,v)\in I}\left(b^u y_b^v \right) \qquad \text{if} \ (i,j)\in I \cr
\delta_{i,i'}\delta_{j,j'} \  \qquad \text{if} \ (i,j)\notin I \cr
\end{array}\right.
\eeq
and the one-form
\beq
\omega^{(i',j')} =  \frac{dx}{P_y(x,y)}\ \sum_{(i,j)\in \Nrond(P)} Q^{(i',j')}_{i,j} \ x^{i-1} y^{j-1}
\eeq
belongs to $\mathcal O^1(\curve)$ and 
\beq
\mathcal O^1(\curve) = \text{span} ( \omega^{(i',j')})_{(i',j')\in \bar I}.
\eeq
In particular we have
\beq
\dim \mathcal O^1(\curve) = |\Nrond(P)|-|\bb|.
\eeq
\el
\smallskip

\proof
First we admit (well known) that $|\bb|< |\Nrond(P)|$.
Let $I\subset \Nrond(P)$ such that $|I|=|\bb|$.
Assume that the matrix 
\beq
B_I=(b^i y_b^j)_{(i,j)\in I, b\in \bb}
\eeq
would be non invertible, in particular, there must exist a vecor $(R_{i,j})_{(i,j)\in I}$ such that
\beq
\forall \ b\in \bb \ , \quad R(b,y_b)=\sum_{(i,j)\in I} R_{i,j} b^i y_b^j = 0
\eeq
i.e. the polynomial $R$ annihiliates all pairs $(b,y_b)$.
However, this polynomial with coefficients in $I\subset \Nrond(P)$ is of degrees too small to be in the ideal generated by $P_x\cap  P_y$ mod $P$, which implies that the set of solutions of $R(b,y_b)=0$ must be strictly smaller than $|\bb|$, which is a contradiction.
Therefore, the matrix $B_I$ is invertible, and thus $B$ has rank $|\bb|$.
Let us consider $Q\in \CC^{\Nrond(P)} $, that we decompose as $Q=Q_I\oplus Q_{\bar I}$.
$\mathcal O^1(\curve)$ is the vector space of $Q \ \frac{dx}{P_y(x,y)}$ such that 
\beq
0 = B. Q = B_I .Q_I + B_{\bar I}. Q_{\bar I},
\eeq
since $B_I$ is invertible, this implies
\beq
Q_I = - B_I^{-1} . B_{\bar I} . Q_{\bar I}.
\eeq
In other words each choice of $Q_{\bar I}$, i.e. $(i',j')\in \bar I$ yields an element of $\mathcal O^1(\curve)$.

The rest of the lemma follows by writing explicitly the inverse matrix in terms of minors.
\eproof

\subsection{Case of other fields}
\label{approoflemmabb}

Consider the case where $P\in \FC[x,y]$ with $\FC$ a commutative field.

\bl\label{lemmabbFC}
all coefficients $Q^{(i',j')}_{i,j}$ defined in  \eqref{defQijlemma} belong to the same field $\FC$ as the coefficients of $P$.
\el

\proof

Notice that the matrix $B_I^{-1} . B_{\bar I}$ belongs a priori to $\FC[\bb]^{I\times \bar I}$, however, the matrix product amounts to a symmetric sum on $b\in \bb$, and thus the matrix elements of $B_I^{-1} . B_{\bar I}$ are symmetric polynomials of the elements of $\bb$, and from lemma \ref{lemsymbpFC} they belong to $\FC$:
\beq
B_I^{-1} . B_{\bar I}\in \FC^{I\times \bar I}.
\eeq
\eproof

\section{Proof of lemma \ref{lemsymbpFC}}
\label{appprooflemmabsym}

\proof
If $b$ is a double point (assumed regular), it must be a  root of $D(x)$, and it thus belongs to an algebraic extension $\FC[b]$.
Moreover $b$ must be a common root of $D(x)$ and $\td D(x)$, therefore the vector $(b,b^2,b^3,b^4,\dots,b^{\deg D+\deg \td D})$ must be in the kernel of the matrix
\beq
R=\left(\begin{array}{ccccccccccc}
D_0 & D_1 & D_2 & \dots & D_{\deg D} & & & & & & \cr
 &  D_0 & D_1 & D_2 & \dots & D_{\deg D} & & & & &  \cr
 & & \ddots & & & & \ddots  & & & & \cr
 & & & \ddots & & & & \ddots  & & &  \cr
 & & & & \ddots & & & & \ddots  & &   \cr
 & & & & & \ddots & & & &\ddots  & \cr
 & & & & & & D_0 & D_1 & D_2 & \dots & D_{\deg D}  \cr
\td D_0 & \td D_1 & \td D_2 & \dots & & & & \td D_{\deg \td D} & & &  \cr
& \td D_0 & \td D_1 & \td D_2 & \dots & & & & \td D_{\deg \td D} & &  \cr
& & \ddots & & & & & & & \ddots &  \cr
& & & \td D_0 & \td D_1 & \td D_2 & \dots & & & & \td D_{\deg \td D}  \cr
\end{array}\right)
\eeq
That matrix must have a kernel of dimension $|\bb|$, and its rank is $\deg D+\deg \td D-|\bb|$.
Let us decompose it by blocks as
\beq
R=\begin{pmatrix}
A & B \cr C & D 
\end{pmatrix}
\eeq
where $D$ is invertible, and all matrices have their coefficients in $\FC$.
Writing that $R.(b,b^2,\dots,b^{\deg D+\deg \td D})=0$ implies that 
\beq
b^{|\bb|+1} = - \sum_{i=1}^{|\bb|} (D^{-1}C)_{|\bb|+1,i} \ b^i
\eeq
This implies that there is a polynomial $S(x)\in \FC[x]$ of degree $|\bb|$ with all coefficients in $\FC$, such that
\beq
\sum_{i=0}^{|\bb|} S_i b^i = 0
\qquad S_{|\bb|}=1
\ , \ S_{i-1} = (D^{-1}C)_{|\bb|+1,i} \ .
\eeq
This implies that integer symmetric polynomials of $b_1,\dots,b_{|\bb|}$ are algebraic combinations of the $S_i$s and thus belong to $\FC$.

Moreover, the symmetric polynomials of branchpoints together with double points, i.e. all the zeros of $D(x)\in \FC[x]$, are in $\FC$.
By subtracting those of the double points, we see that integer symmetric polynomials of branchpoints are in $\FC$.
\eproof

Remark that the proof needs to compute the inverse $D^{-1}$, and in particular needs to divide by $\det D$.
This is where we needed a field rather than a ring, and also we simplified the discussion by assuming characteristic 0.

\end{document}